\documentstyle[11pt,aaspp4]{article}
\def\plotfiddle#1#2#3#4#5#6#7{\centering \leavevmode
    \vbox to#2{\rule{0pt}{#2}}
    \includegraphics{#1}}
\newcommand{\Mdot}{M$_{\odot}$}
\newcommand{\msun}{M$_{\odot}$}
\newcommand{\msunyr}{M$_{\odot}\,$yr$^{-1}$}
\newcommand{\msunpc}{M$_{\odot}\,$pc$^{-3}$}
\newcommand{\ergs}{erg s$^{-1}$}
\newcommand{\gcc}{g$\,$cm$^{-3}$}
\newcommand{\kms}{km$\,$s$^{-1}$}
\newcommand{\degrees}{$^{\circ}$}
\renewcommand{\deg}{$^{\circ}$}
\newcommand{\hto}{H$_2$O}
\renewcommand{\_}{$\,$}
\newcommand{\sol}{$_{\odot}$}
\newcommand{\sm}{\scriptsize}

\newlength{\phantomdigit}
\newcommand{\z}{\hspace*{\phantomdigit}}

\begin{document}
\thispagestyle{empty}
\begin{center}

OBSERVATIONAL EVIDENCE FOR MASSIVE BLACK HOLES\\
IN THE CENTERS OF ACTIVE GALAXIES

J. M. Moran, L. J. Greenhill 

Harvard-Smithsonian Center for Astrophysics 

and J. R. Herrnstein

National Radio Astronomy Observatory
\end{center}

\begin{abstract}

Naturally occurring water vapor maser emission at 1.35 cm wavelength provides an accurate
probe for the study of accretion disks around highly compact objects, thought
to be black holes, in the centers of active galaxies. Because of the exceptionally fine
angular resolution, 200 microarcseconds, obtainable with very long baseline interferometry,
accompanied by high spectral resolution, $<0.1$~\kms, the dynamics and structures of
these disks can be probed with exceptional clarity. The data on the galaxy NGC\_4258 
are discussed here in detail. The mass of the black hole binding the accretion disk 
is $3.9 \times 10^7$~\msun. 
Although the accretion disk has a rotational period of about 800 years, the physical
motions of the masers have been directly measured with VLBI over a period of a few years. 
These measurements also allow
the distance from the earth to the black hole to be estimated to an accuracy of 4 percent. 
The status of the search for other maser/black hole candidates is also discussed.

\end{abstract}

\section{INTRODUCTION}

The observational evidence for the existence of supermassive
black holes ($10^6$--$10^9$ times the mass of the sun, \msun) 
in the centers of active galaxies
has been accumulating at an ever accelerating pace for the last 
few decades (e.g., Rees 1998; Blandford \& Gehrels 1999). Seyfert (1943) first drew attention to
a group of galaxies with unusual excitation conditions
in their nuclei indicative of energetic activity. Among the twelve 
galaxies in his list
was NGC\_4258, which is the subject of much of this paper. 
Such galaxies, now known as galaxies with active galactic nuclei (AGN),
have grown in membership and importance. Ironically, NGC\_4258  no longer  belongs
to the class of Seyfert galaxies  by modern classification standards 
(Heckman 1980), but it is still
considered to have a mildly active galactic nucleus. 
Meanwhile, the study of AGN has become a major field in modern
astrophysics. In the 1960s, galaxies with AGN were discovered with
intense radio emission arising from  jets of relativistic
particles often extending far beyond the optical boundaries of the 
host galaxy. The central engine, the source of energy that powers
such jets and other phenomena in the centers of galaxies, has long 
been ascribed to black holes (e.g., Salpeter 1964; Blandford \& Rees 1992). 
There are two sources of energy for these phenomena: the
gravitational energy from material falling onto the black hole and the spin energy of the
black hole itself (Blandford \& Znajek 1977).

The direct evidence for black holes in AGN has come principally from
observations of the motions of gas and stars in the  extended environments
of black holes. In the optical and infrared domains, the evidence
for black holes from stellar measurements comes from an analysis
of the velocity dispersion of stars as a function of distance from
the dynamical centers of galaxies. In the case of our own Galactic center, the 
proper motions (angular velocities in the plane of the sky)
of individual stars can be measured.
These data show that there is a mass of about $2.6 \times 10^6$~\Mdot\ within a volume
of radius 0.01 pc (Genzel et al. 1997; Ghez et al. 1998).
In addition, measurements
by the Hubble Space Telescope of the velocity field of hydrogen gas in
active galaxies indicate the presence of  massive centrally condensed objects. 
Reviews of these data have been written by Faber (1999), Ho (1999), 
Kormendy and Richstone (1995), and others. 

In the X-ray portion of
the spectrum, there is compelling evidence for black
holes in AGN from the detection of the highly broadened iron K$\alpha$ line at 6.4 keV.
The line
is broadened by the gravitational redshift of gas as close as 3
Schwarzschild radii from the black hole. An example of an 
iron line profile in the galaxy MCG\_-6-30-15 is shown in Figure~1 (Tanaka et al. 1995).
The linear
extent of the emission region cannot be determined directly by the X-ray
telescope, so it is not possible to estimate directly the mass of the
putative black hole. Detailed analysis of the line profile suggests that the black 
hole is spinning
(e.g., Bromley, Miller, \& Pariev 1998).

\begin{figure}[t]
\plotfiddle{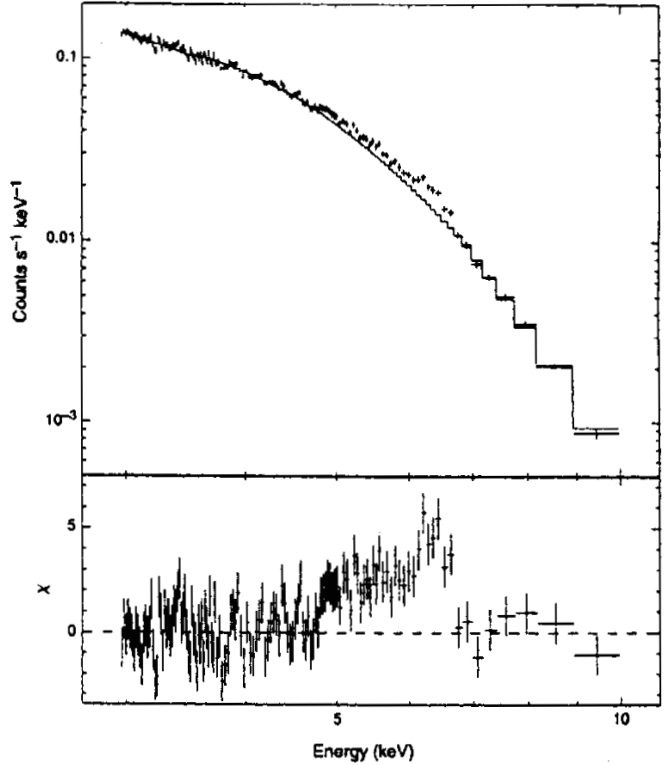}{3.25in}{.2}{90}{90}{-275}{-204}
\caption{\setlength{\baselineskip}{8 pt} The X-ray spectrum of the galaxy MCG\_-6-30-15, observed
by the Japanese ASCA satellite. The top panel shows the total spectrum with a 
model of the continuum emission fitted to the data outside the range of 5--7 
keV. The bottom panel shows the residuals, which reveal a broad spectral 
feature attributed to the Fe K$\alpha$ line at 6.4 keV. The line has a width of
100,000 \kms. The most extreme redshifted part is thought to
arise from gas at a radius of about 3 Schwarzschild radii. (From Tanaka et al. 1995)}
\end{figure}

In the radio regime, a new line of inquiry
has given unexpectedly clear and compelling evidence for 
black holes: the discovery of water masers orbiting highly massive
and compact central objects. With the aid of very long baseline
interferometry (VLBI), which provides angular resolution as fine as 200~microarcseconds
($\mu$as) at a wavelength of 1.3~cm
and spectral resolution of 0.1 km$\,$s$^{-1}$ or less,
the structure of accreting material around 
these central objects can be studied in detail. This paper describes 
the observations and the significance of these measurements of water masers in AGN. 
We begin with a brief description of cosmic masers and the interferometric
techniques used to observe them.

\section{COSMIC MASERS}

Intense maser action in cosmic molecular clouds was 
discovered in 1965 (Weaver et al. 1965) from observations of OH, and later from \hto, SiO, and 
CH$_3$OH. In the case of water vapor, the commonly observed masers
emit in the 6$_{16}$--5$_{23}$ transition at 22235 MHz (1.35 cm wavelength).
Most masers
have been found to be associated with one of two types of objects: newly
formed stars or evolved stars (e.g., Elitzur 1992; Reid \& Moran 1988).  Although 
very distinct, they share the characteristic of having envelopes of 
outflowing gas and dust (silicate material). 
The pump source in all cases is thought to come in the form
of either shock waves or infrared radiation.
More recently, masers have been found in the spiral
arms of nearby galaxies and in AGN.

Cosmic masers are similar to their laboratory counterparts on earth
in that their intense radiation is produced by population inversion. However,
cosmic masers are one-pass amplifiers and have little temporal or spatial
coherence. The intensity of cosmic masers varies, often erratically, on 
timescales of hours to years. The underlying electric field is
a Gaussian random process (Moran 1981). In an unsaturated maser (in astronomical 
terminology), the pumping is sufficiently strong that the microwave intensity 
does not affect the level populations, and the intensity increases exponentially
through the masing medium. The input signal can be either a 
background source or the maser's own spontaneous emission. In a saturated maser,
one pump photon is needed for each microwave photon, and in a one-dimensional
maser medium where beaming can be neglected, the intensity increases linearly
with distance.
Maser emission is expected to be beamed. Most masers are thought to be saturated,
and this condition requires the least pump power. However, this assertion
is difficult to prove observationally.

\begin{figure}[t]
\plotfiddle{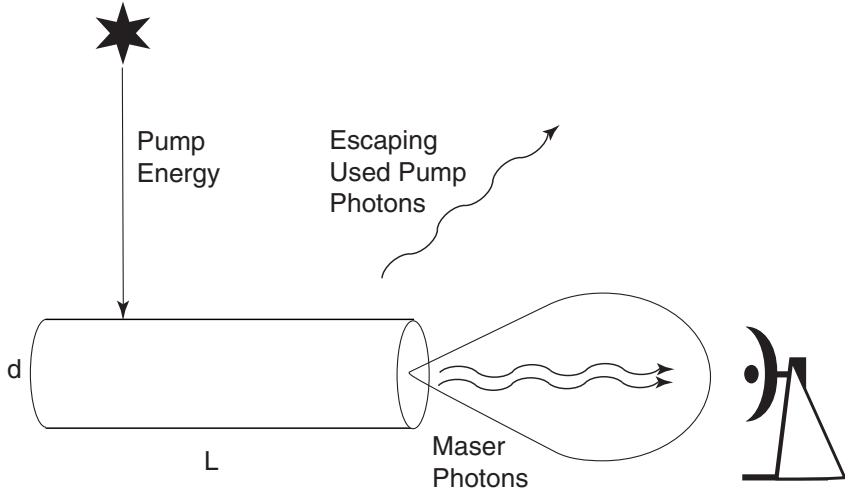}{1.95in}{00}{55}{55}{-185}{-140}
\caption{\setlength{\baselineskip}{8 pt} A cartoon of a simple filamentary maser.
The pump energy
can be supplied by either shock waves or radiation. Pump cycles usually 
involve excitation to infrared rotational levels, followed by de-excitation
to the upper level of the maser transition. In order to avoid thermalization,
the infrared photons emitted during the pump cycle must escape from the
masing medium, which favors a geometry that is thin in at least one dimension.
In a filamentary geometry, the masing medium is defined 
either by the physical extent of the masing gas or by the volume over which 
the gas is coherent, that is, where the variation in the velocity projected along the 
line of sight is less than the thermal line width. The maser emission is
beamed into a cone of angular opening equal to $d/L$. If the maser amplifies
a background source, the radiation will be beamed in the forward direction.
If it amplifies its own spontaneous emission, then beamed maser emission will
emerge from both ends, and weaker emission from the sidewalls.}
\end{figure}

Consider a simple geometry for a maser, a filamentary tube, shown in 
Figure~2. The boundaries
of the filament can be defined in terms of either gas density or the region where
the line-of-sight velocity is constant to within the thermal line width. The
gas in the tube is predominantly molecular hydrogen, with trace amounts
of water vapor (about one part in $10^{5}$) and other constituents. If this
maser medium is saturated, then the luminosity is given by
\begin{equation}
L  = h \nu n\Delta P V~,
\end{equation}
where $h$ is Planck's constant, $\nu$ is the frequency,
$n$ is the population density in pumping level,
$\Delta P$ is the differential pump rate per molecule,
and $V$ is the volume of the masing cloud.  
This is the most luminosity a maser of given pump rate and volume can produce.
The emission will be beamed along the major axis of the filament into
an angle  
\begin{equation}
\beta \cong {d\over L}~,
\end{equation}
where $d$ is the cross-sectional diameter, and $L$ is the length of the filament.
The beam angle and the length
of the maser are not directly observable. Since the volume of the
maser is approximately $d^2L$, the observed flux density from a maser 
beamed toward the earth is 
\begin{equation}
F_{\nu} = {1\over 2} h \nu n{{\Delta P}\over {\Delta \nu}} {V\over {D^2\Omega}}%
  = {1\over 2} h \nu n{{\Delta P}\over {\Delta \nu}} {L^3\over D^2}~,
\end{equation}

\noindent
where $\Delta \nu$ is the line width, $D$
is the distance between the maser and the observer,
and $\Omega$ is the beam angle of the emission, $\sim \beta^2$. 
The maximum allowable hydrogen number density is about $10^{10}$ molecules 
per cubic centimeter, above which the maser levels become thermalized by collisions. This
maximum allowable 
density ($\rho_c = 3 \times 10^{-13}$~\gcc, an important parameter in much of the 
following discussion) along with the maximum allowable pump rate, which equals the
Einstein A-coefficient for infrared transitions linking the maser 
levels, limit the luminosity of a maser of given volume.

Water masers outside our Galaxy were first discovered in the spiral arms of
the nearby galaxy M33 by Churchwell et al. (1977). Their properties were found to 
be similar to masers found in Galactic star-forming regions. Much more 
luminous water masers were found in the AGN associated with NGC\_4945 and the Circinus
galaxy by dos Santos \& Lepine (1979) and Gardner \& Whiteoak (1982), respectively.
Claussen, Heiligman, \& Lo (1984) and Claussen \& Lo
(1986) conducted surveys and found five additional masers associated with AGN, including
the one in NGC\_4258 (see Figure~3). They suggested that these masers might arise
in gas associated with dust-laden molecular tori that had been proposed to
surround black holes by Antonucci \& Miller (1985). Nakai, Inoue, \& Miyoshi (1993), with a
powerful new spectrometer of 16,000 channels spanning a velocity range
of 3000 \kms, observed NGC\_4258 and discovered satellite line clusters
offset from the systemic velocity by about $\pm$ 1000 \kms, which are shown in
Figure~3 (see also Miyoshi 1999).

\begin{figure}[t]
\plotfiddle{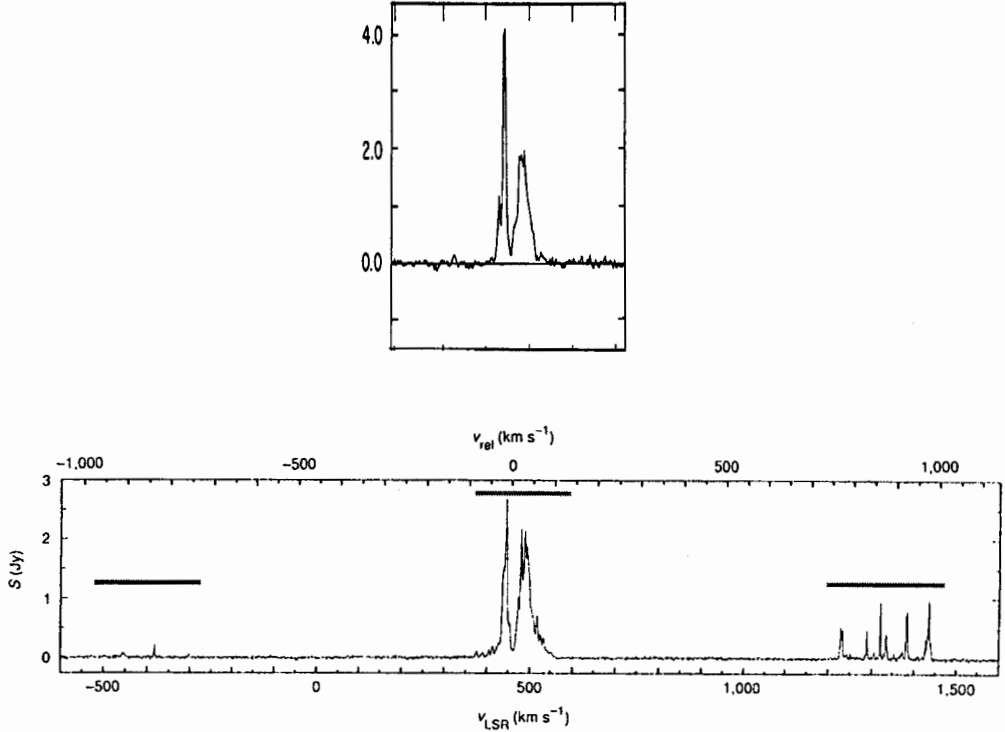}{3.2in}{0.3}{100}{100}{-295}{-253}
\caption{\setlength{\baselineskip}{8 pt}
The top panel shows the maser spectrum as first discovered in NGC\_4258 
at velocities near the systemic velocity of the galaxy (Claussen, Heiligman, \& Lo 1984).
The lower spectrum shows the observations of Nakai, Inoue, \& Miyoshi (1993) over
a much broader velocity range. The velocities are computed from the Doppler
effect and are based on a rest frequency of 22235.080 MHz. The effects of 
the motions of the earth and sun with respect to the local standard of rest
have been removed. The bars indicate velocity ranges of emission.}
\end{figure}

The compelling reason that the radio emission from the water vapor transition
arises from the maser process is straightforward. A typical example of a maser
line from a small part of the spectrum of NGC\_4258 is shown in
Figure~4. The line width is about 1 \kms, the thermal broadening
expected for a gas cloud at 300~K. However, the angular size determined by
radio interferometry is
less than 100 $\mu$as, implying that the equivalent blackbody 
temperature must exceed $10^{14}$~K. The exceedingly high brightness
of the radiation is the principal evidence for the maser process.
(Typical molecular lines from molecular clouds have velocities of 
several tens of \kms\ --- due to thermal and turbulent broadening --- and
brightness temperatures of less than 100~K.)
Cosmic masers produce very bright spots of radiation but have
little else in common with terrestrial masers.
It is difficult to use masers to determine physical conditions (e.g.,
temperature, density) in
molecular clouds because of the complexity of the maser process. However,
as compact sources of narrowband radiation, masers are ideal probes of the
dynamics of their environment.

\begin{figure}[t]
\plotfiddle{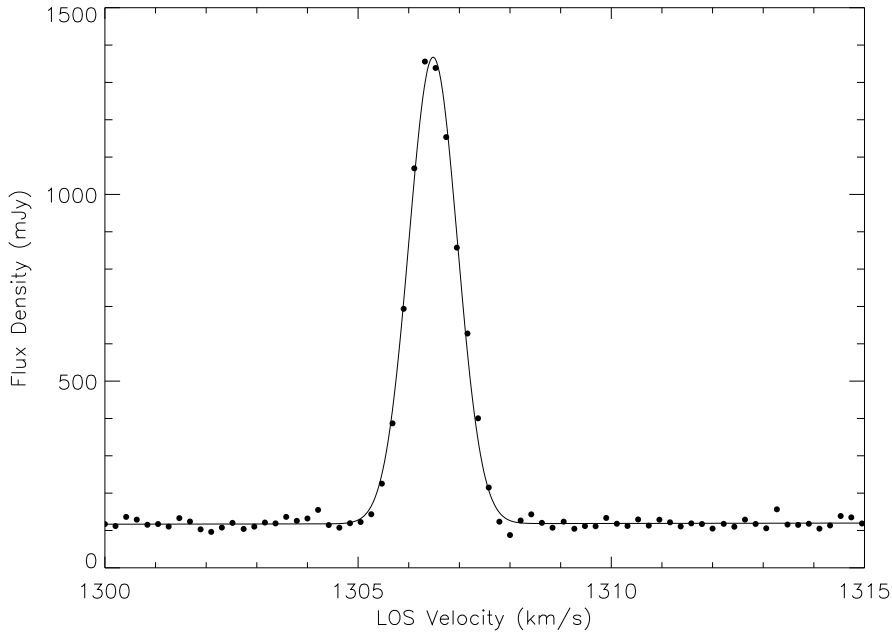}{2.7in}{0}{68}{68}{-215}{-125}
\caption{\setlength{\baselineskip}{8 pt}
An expanded view of the spectral feature near 1306 \kms\ in NGC\_4258, which is typical
of emission from an isolated maser component (see Figure~3). The line width
is characteristic of gas at 300~K, but the intensity corresponds to that
of a blackbody at an equivalent temperature greater than 10$^{14}$~K.}
\end{figure}

\section{VLBI}

\begin{figure}[t]
\plotfiddle{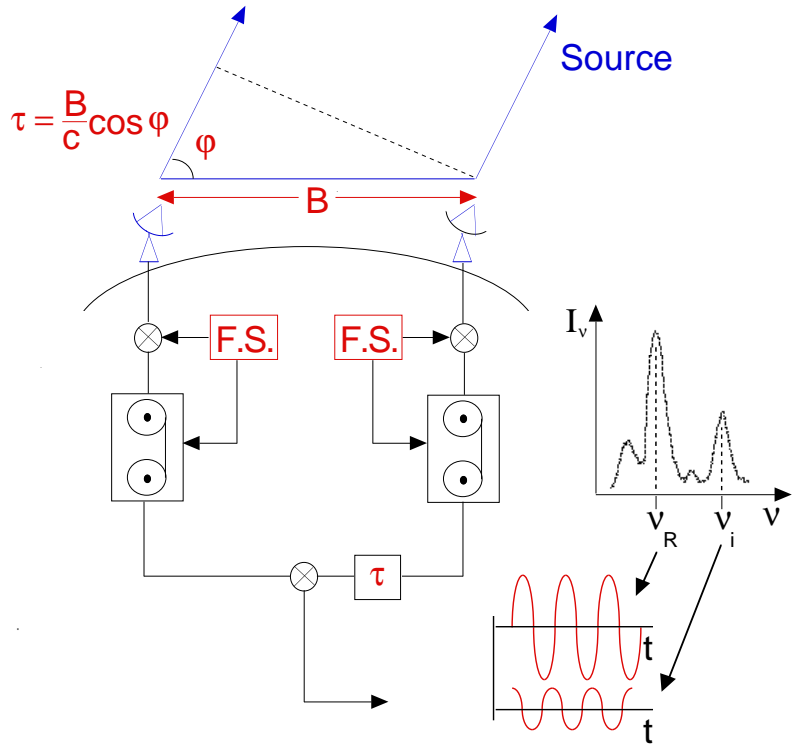}{3.25in}{00}{55}{55}{-168}{-115}
\caption{\setlength{\baselineskip}{8 pt}
A block diagram of a two-element very long baseline interferometer.
It operates as a coherent interferometer.  An atomic
frequency standard (F.S.) controls the phase of the local oscillator
signal at each telescope used to convert the radio frequency signal
to a video band for  recording  on magnetic tape 
in digital form (without square-law detection) and sampled at the
appropriate Nyquist rate. On playback, one
of the signals is delayed by $\tau$ to compensate for the differential
propagation time from the source to the antennas. The signals are 
correlated and Fourier transformed
to produce cross-power spectra, or correlated power as a function
of frequency and time. }
\end{figure}

The most important tool for the study of the angular structure of masers
is very long baseline interferometry (VLBI). Signals from a maser,
or from other bright compact radio sources, are converted to a low-frequency
baseband and recorded in digital format on magnetic tape at Nyquist sampling
rates of up to about
$10^8$ samples per second at widely separated telescopes that
operate independently. They form a radio version of the classical  
Michelson stellar interferometer,
whose coherence is maintained by the use of atomic frequency standards
to preserve the signal phase and timing (see Figure~5). The received signals
(which are proportional to the incident electric fields)
from an array of two or more telescopes are cross-correlated pairwise
to form cross-correlation functions. Taking advantage of the earth's 
rotation, the spatial  
cross-correlation function of the incident electromagnetic field, or visibility,
can be measured
over a wide range of projected baseline vectors. The image and fringe
visibility functions are related through a Fourier transform (see Thompson,
Moran, \& Swenson 1986). The
temporal Fourier transform of the cross-correlation function gives the
cross-power spectrum of the radiation, or visibility as a function of frequency, 
so that images at different frequencies can be obtained.
The intrinsic angular resolution, $\theta$, 
of a multielement interferometer is $0.7 \lambda/B$, where $\lambda$ 
is the wavelength, and $B$ is the longest
baseline length. For water vapor, $\lambda$ = 1.35 cm, and $B$ is typically 6000~km,
which gives a resolution of 200~$\mu$as. The spectral resolution
available is typically about 15~KHz, or about a fifth of the line widths (see Figure~4). 
In maser sources, one spectral feature at a particular frequency or velocity can
be used as a phase reference for the interferometer, and all other
phases referred to it. With this technique, the coherence time of the
interferometer can be extended indefinitely, and the relative positions
of masers with respect to the reference feature can be measured to a small
fraction of the fringe spacing, or intrinsic resolution. The relative 
position of an unresolved maser component can be measured to an
accuracy of about
\begin{equation}
\Delta \theta  = {1\over 2} {\theta \over {{S\!N\!R}}}~,
\end{equation}
where ${S\!N\!R}$ is the signal-to-noise ratio.

\section{THE STUDY OF NGC\_4258}

The imaging of the maser in NGC\_4258 was one of the first projects undertaken
by a dedicated VLBI system known as the Very Long Baseline 
Array (VLBA) (see Figure~6) in the spring of 1994. 
Previous VLBI measurements of the systemic features had shown that
they  arose from an elongated structure with a velocity
gradient along the major axis, highly suggestive of a rotating disk seen edge-on
(Greenhill et al. 1995a). The VLBA measurements of all the maser components provide convincing
evidence for a rotating disk around a  massive central
object. 

\begin{figure}[t]
\plotfiddle{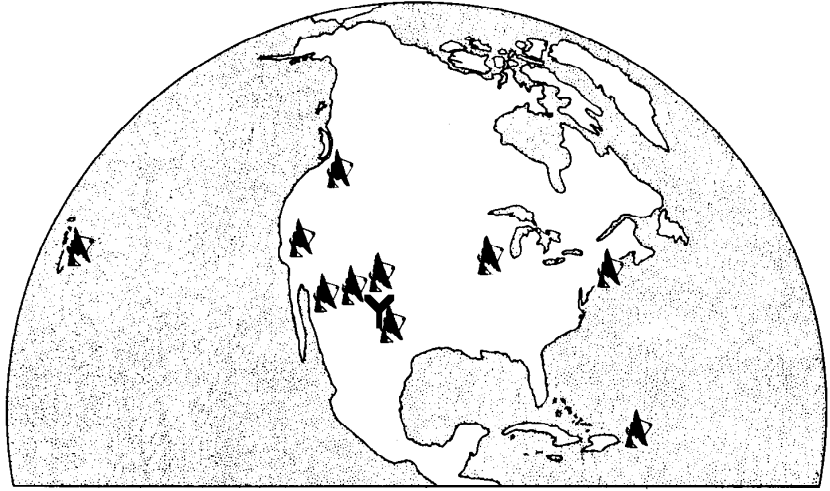}{2.05in}{00}{75}{75}{-228}{-174}
\caption{\setlength{\baselineskip}{8 pt}
The distribution of the ten elements of the Very Long Baseline
Array (VLBA). This network is often augmented with other radio telescopes
such as the Very Large Array (a 27-element array operating as a 
phased array), shown here with the symbol {\sf Y}, and the 100 m telescope of the Max Planck Institute for 
Radio Astronomy near Bonn, Germany.}
\end{figure}

\begin{table}[th]
\begin{center}
\begin{tabular}{l@{\hspace{1in}}l}
\multicolumn{2}{c}{TABLE 1}\\[1ex]
\multicolumn{2}{c}{\sc Parameters of Molecular Disk Traced by Water Vapor Masers %
                       in NGC$\,4258^a$}\\[1ex]
\hline\hline\noalign{\vspace{1ex}}
Inner radius, $R_i$ 				& 0.14 pc (3.9 mas)  \\
Outer radius, $R_o$				& 0.28 pc (8.0 mas)  \\
Inner rotation velocity, $v_{\phi}$($R_i$) 	& 1100 \kms \\
Outer rotation velocity, $v_{\phi}$($R_o$)	& 770 \kms \\
Inner rotation period				& 800 yrs \\
Outer rotation period				& 2200 yrs \\
Position angle of disk (at 3.9 mas radius)	& 80\deg \\
Inclination angle				& 98\deg \\
Position velocity slope				& 282 \kms $\,$mas$^{-1}$ \\
Central mass, $M$				& $3.9 \times 10^7$ M\sol \\
Disk mass, $M_d$				& $ < 10^6$ M\sol \\
Central mass density, uniform distribution	& $> 4 \times 10^9$ M\sol $\,$pc$^{-3}$ \\
Central mass, Plummer distribution		& $ > 10^{12}$ M\sol $\,$pc$^{-3}$ \\
Centripetal acceleration, systemic features	& $9.3$ \kms $\,$yr$^{-1}$ \\
Centripetal acceleration, high-velocity	features & $\le 0.8$ \kms $\,$yr$^{-1}$ \\
Disk systemic velocity$^b$, $v_0$		& 476 \kms \\
Galactic systemic velocity (optical)$^{b,c}$	& 472 \kms \\
Radial drift velocity, $v_R$     		& $< 10$ \kms \\
Thickness of disk, $H$				& $< 0.0003$ pc \\
Maser beam angle, $\beta$			& 8\deg \\
Disk--galaxy angle$^d$				& 119\deg \\
Apparent maser luminosity			& 150 L\sol \\
Model luminosity$^e$				& 11 L\sol \\
Distance, $D$					& $7.2\pm 0.3$ Mpc \\
\noalign{\vspace{.5ex}}\hline \noalign{\vspace{1.5ex}}
\multicolumn{2}{l}{$^a$Based on the distance estimate of 7.2 Mpc.} \\
\multicolumn{2}{l}{$^b$\vtop{\raggedright
				Radio definition, with respect to the local standard of rest.
				To convert to heliocentric velocity (radio), subtract 8.2 \kms;
				to convert to heliocentric (optical), subtract 7.5 \kms.}} \\
\multicolumn{2}{l}{$^c$From Cecil, Wilson, \& Tully (1992).} \\
\multicolumn{2}{l}{$^d$Angle between the spin axis of the molecular disk and the spin
			axis of the galaxy.} \\
\multicolumn{2}{l}{$^e$Radiation into a zone within $\pm 4$\deg of the plane of the disk.} \\

\end{tabular}
\end{center}
\end{table}

The basic observational results on NGC\_4258 obtained over the past few years 
can be summarized as follows (see Table 1 for a list of parameters):

\begin{enumerate}
\item The masers appear to trace a highly elongated, although slightly curved,
      structure (Figure~7). The high-velocity, redshifted and blueshifted features
      are offset in position on the left and right sides of the systemic features,
      respectively. The velocities of the high-velocity features as a function of
      impact parameter (position along the major axis of the distribution) follow
      the prediction of Kepler's third law of orbital motion. The systemic features
      show linear dependence with impact parameter (Miyoshi et al. 1995).
\pagebreak

\item The distribution of maser features in the direction normal
      to the major axis is too small to be measured at present (see Figure~7).
      The upper limit
      on the ratio of thickness to radius of the disk is 0.0025 
      (Moran et al. 1995). 

\item The upper limit of any toroidal component of the magnetic field in the
      masers, derived
      from searches for Zeeman splitting in the line at 1306 \kms, is less than 
      300 mG (Herrnstein et al. 1998a).

\item The accelerations (i.e., the linear drift in the line-of-sight velocity with time) 
      of the systemic features are about 9 \kms$\,$yr$^{-1}$
      (Haschick, Baan, \& Peng 1994; Greenhill et al. 1995b; Nakai et al. 1995). 
      The high-velocity features that have been tracked have accelerations in the range $\pm 0.8$
      \kms$\,$yr$^{-1}$ (Bragg et al. 1999). 

\item The high-velocity features show no proper motions with respect to a 
      fixed-velocity component in the systemic range (Herrnstein 1996). The systemic features
      show proper motions of about \mbox{32 $\mu$as\_yr$^{-1}$} (Herrnstein et al. 1999).

\item There is an elongated continuum radio source, which appears to be a jet 
      emanating from the black hole position, parallel to the axis of rotation
      (Herrnstein et al. 1997). There is no 1.35~cm wavelength emission from the
      position of the black hole (Herrnstein et al. 1998b).       
\end{enumerate}

\begin{figure}[t]
\plotfiddle{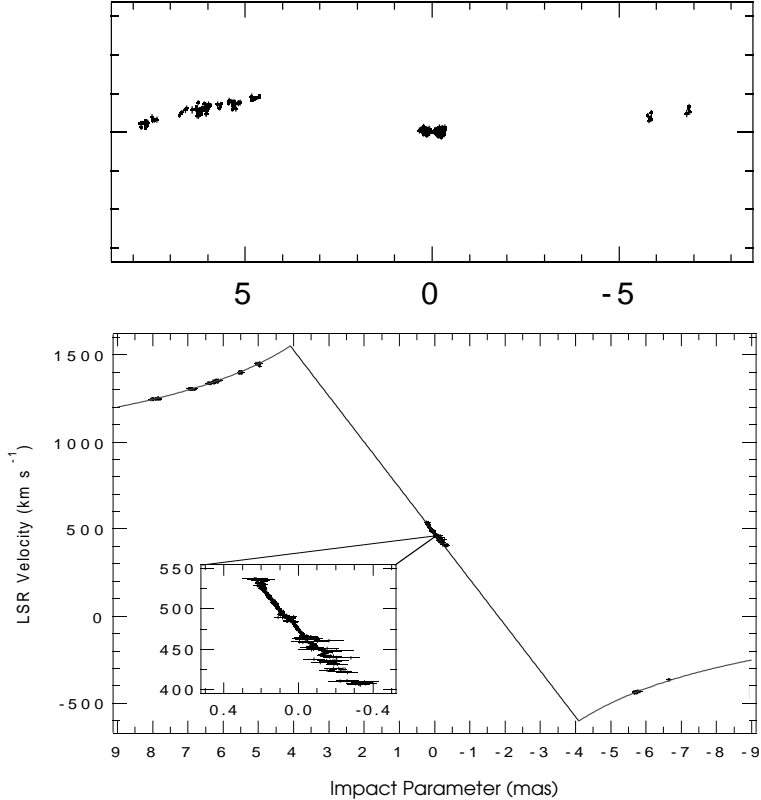}{3.7in}{00}{90}{90}{-275}{-185}
\caption{\setlength{\baselineskip}{8 pt}
Top: Image of the maser emission from the nucleus of NGC\_4258. The
ticks on the axes are in milliarcseconds. One milliarcsecond corresponds to
0.035 pc, or $1.1 \times 10^{17}$~cm, at a distance of 7.2~Mpc.
Bottom: The line-of-sight
velocities of the masers versus position along the major axis. The curved
portions of the plot precisely follow a Keplerian dependence. Data from
January 1995 (top) and April 1994 (bottom).}
\end{figure}

There is virtually no doubt that the masers trace a very thin disk in
nearly perfect Keplerian motion. Five of six phase-space parameters
have been measured for each maser spot, two spatial coordinates and
three velocity coordinates. The missing
coordinate is the position along the line of sight, which must be 
inferred from the constraint provided by Kepler's third law.

The approximate placement of the masers in the  disk can be understood 
by considering
a simple thin, flat disk viewed edge-on. In this case the
line-of-sight velocity, $v_z$,  of a maser will be given by
\begin{equation}
v_z - v_0 = \sqrt{{GM}\over R} \sin \phi~,
\end{equation}
where $v_0$ is the line-of-sight velocity of the central object (i.e., the
systemic velocity), $G$ is the gravitational constant, $R$ is the distance
of a maser component from the black hole,
and $\phi$ is the azimuth angle in the disk, measured from the line between
the black hole and the observer. If the disk were randomly 
filled with observable masers, one might 
expect to see a velocity position diagram as shown in Figure~8.
The linear boundaries of the distribution are populated by masers at 
the inner and outer edges of the annular disk. The masers on the curved boundaries 
lie on the midline, where $\phi$ = 90\degrees . Hence, the masers in NGC\_4258 have a 
very specific distribution: the high-velocity masers lie close to the
midline, and the systemic
masers lie within a narrow range of radii. From Equation 5, the radius of a
particular maser can be determined as
\begin{equation}
R = {(GM)}^{1\over 3} {\left[{b}\over {v_z-v_0}\right]}^{2\over 3}~,
\end{equation}
where ${b}$ is the projected distance on the sky along the major axis from the center of the
disk ($\sin \phi = b/R$). Similarly, positional offsets from the midline,
$z$, of the high-velocity features can be determined by deviations from a Keplerian
curve; that is, 
\begin{equation}
z = \sqrt{R^2 - b^2}~.
\end{equation}
There is a two-fold ambiguity in the $z$ component of the position for the
edge-on disk case. Unambiguous estimates of the positions of the high-velocity 
features have been derived
from the accelerations by Bragg et al. (1999), who showed that the masers
lie within 15 degrees of the midline.  

\begin{figure}[t]
\plotfiddle{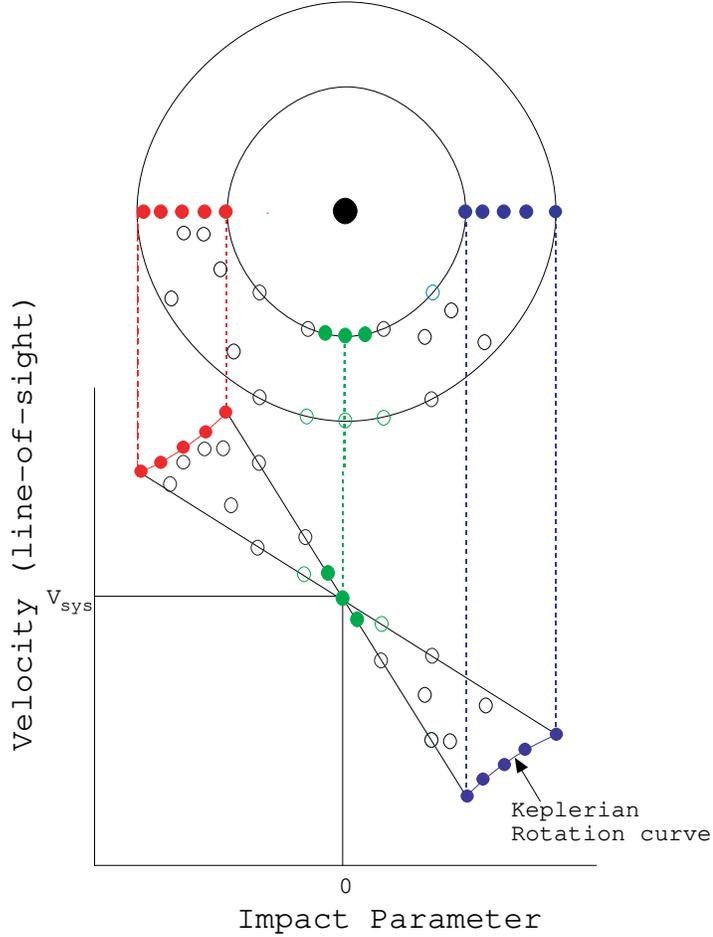}{4.2in}{00}{50}{50}{-140}{-5}
\caption{\setlength{\baselineskip}{8 pt}
Top: A cartoon model of a flat annular disk viewed edge-on, with randomly
distributed maser sources. Bottom: Each maser will appear as a point within
the ``bow tie'' boundary in the plot of line-of-sight velocity versus impact 
parameter. The curved portion of the boundary is populated by masers
located along the midline, the diameter perpendicular to the line of sight. 
This is a pure Keplerian curve, because the velocity vectors lie
along the line of sight. Masers at a fixed radius will appear 
along a straight line. The steep and shallow lines correspond to
masers on the inner and outer annular boundaries of the disk.}
\end{figure}

An expanded plot of the Keplerian part of the velocity curve is
shown in Figure~9. The data fit a Keplerian curve to an accuracy
of about 3 \kms, or less than 1 percent of the rotation speed. However,
there are noticeable deviations from a perfect fit. The estimate of the 
central mass of the disk derived from this data depends on the distance 
to the maser, and has a value  of $3.9 \times 10^7$ \Mdot~for a distance 
of 7.2 Mpc. This mass corresponds to an Eddington luminosity (where
radiation pressure from Thomson scattering would balance gravity) of
$5 \times 10^{45}$~erg s$^{-1}$.   
Since the total electromagnetic emission appears to be less than $10^{42}$~\ergs,
the system is highly sub-Eddington. 

\begin{figure}[t]
\plotfiddle{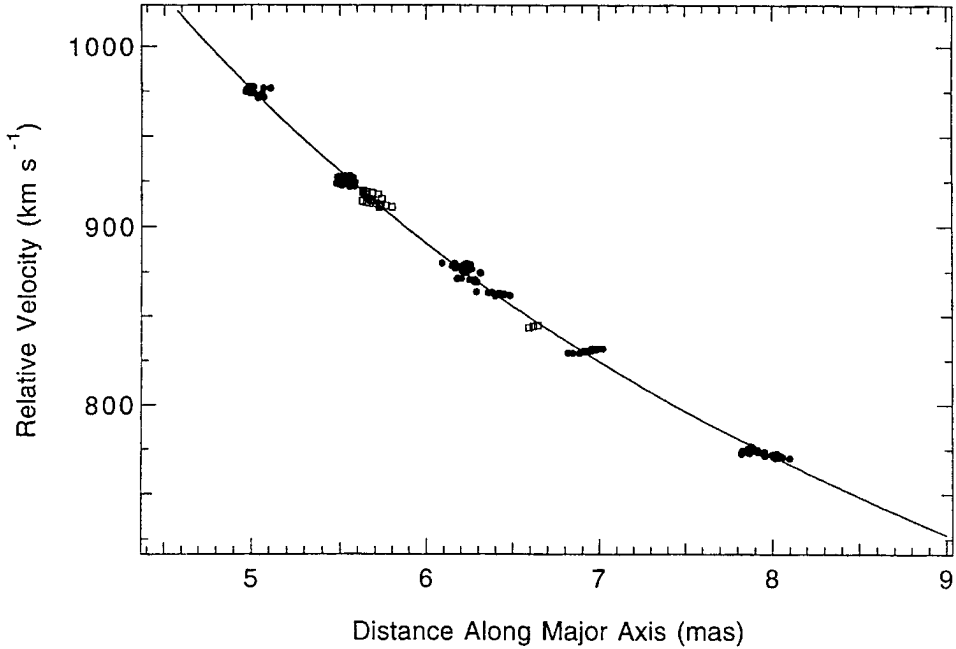}{2.9in}{00.0}{100}{100}{-300}{-260}
\caption{\setlength{\baselineskip}{8 pt}
The magnitude of the line-of-sight velocities of the masers, relative
to the systemic velocity, versus distance from the dynamical center
of NGC\_4258. The filled circles are the redshifted masers, and the squares
are the blueshifted masers. (Data from April 1994)}
\end{figure}


Since this binding mass must lie inside the inner radius of the maser
disk, the mass density, assuming a spherical mass distribution, is
at least $4 \times 10^9$ \msunpc\ ($3 \times 10^{-13}$~\gcc). 
It is unlikely that this mass is in the form
of a dense star cluster (Maoz 1995). The average density of stars in the solar 
neighborhood is about 1~\msunpc, and the density of the densest
known star cluster is about $10^5$~\msunpc. A star cluster will have
a mass distribution that decreases monotonically with radius.
In order not to disrupt the Keplerian
curve, the core mass for a reasonable distribution must have a peak
density of at least $1 \times 10^{12}$~\msunpc. A cluster of massive stars at
this density would evaporate from gravitational interactions on
a timescale short with respect to the age of the galaxy, while a
cluster of low-mass stars would destroy itself from collisions over
a similar timescale. Hence, it is unlikely that the central mass
is in the form of a star cluster (see also Begelman \& Rees 1978). 
The best explanation is that
the central object is a supermassive black hole, with a Schwarzschild
radius ($R_S$) of about $1.2 \times 10^{13} $ cm. Hence, the masers are distributed
in a zone between 40,000 and 80,000~R$_S$. Because the maser clouds are so far
from the event horizon, deviations of their motions from the predictions of Newtonian mechanics
are small.
The gravitational redshift and transverse Doppler shift are about 4~\kms\ (detectable),
the expected Lense-Thirring precession (see below)
is less than about 3\degrees over the maser annulus (possibly detectable), and  
the apparent shift of the maser positions due to gravitational bending is about
0.1 $\mu$as (undetectable). 

The disk is remarkably thin. In a disk supported against gravity by
pressure (hydrostatic equilibrium), the density distribution is expected to have 
a Gaussian profile with a thickness, $H$, given by the relation
\begin{equation}
H/R = {(c_s^2+v_a^2)}^{1\over 2}/v_\phi~,
\end{equation}
where  $c_s$ is the sound speed and $v_a$ is the Alfv\'en speed,
which characterize thermal and  magnetic support pressure, respectively,
and $v_\phi$ is the Keplerian rotational speed.
Since $H/R < 0.0025$,
the quadrature sum of the sound speed and Alfv\'en speed is less than 2.5~km s$^{-1}$. 
The upper limit on the magnetic field of 300 mG suggests that the Alfv\'en speed, 
$B/\sqrt{4\pi\rho}$, where $\rho$ is the density, is less than 3 \kms\ for $\rho=\rho_c$
(the critical density for quenching maser emission). If the support were
completely due to thermal pressure, the temperature would be less than 1000~K. 

A proper determination of the positions of the masers on the disk requires
that the warp and the inclination of the
disk to the line of sight be taken into account.  An example of such a  
disk, slightly warped (in position angle only) and slightly  inclined to the line
of sight, that fits the maser distribution in position and velocity
is shown in Figure~10. 

\begin{figure}[t]
\plotfiddle{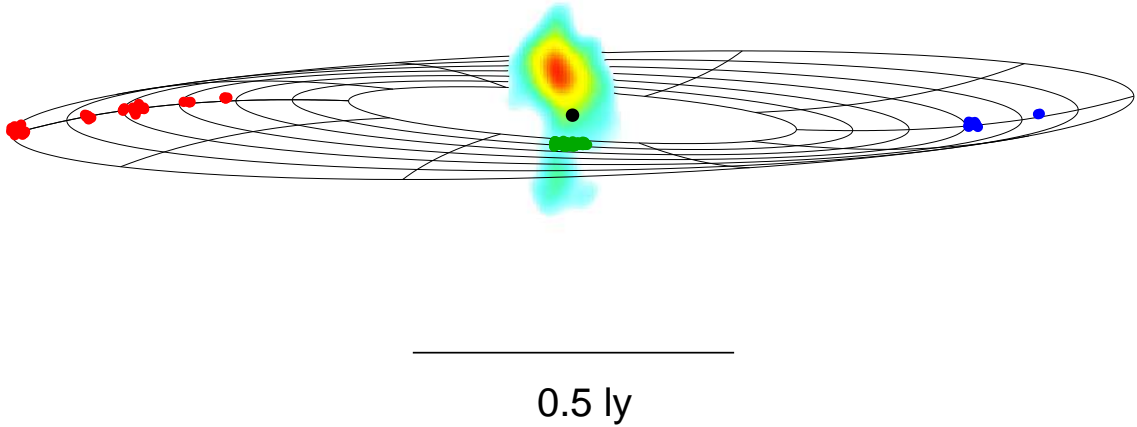}{1.75in}{90}{65}{65}{255}{-80}
\caption{\setlength{\baselineskip}{8 pt}
The warped annular disk (wire mesh) modeled to the maser positions,
velocities, and accelerations (adapted from Herrnstein, Greenhill, \& Moran 1996).
The black dot in the center marks the dynamical center
of the disk. The continuum emission at 1.3 cm is shown in the shaded gray
contours. The southern jet may be weaker than the northern jet because of thermal
absorption in the disk. The lack of emission at the position of the black hole 
places constraints on any coronal or advection zone surrounding the black hole (Herrnstein
et al. 1998b).}
\end{figure}

The distance to the maser of 7.2 $\pm$ 0.3 Mpc was determined from
analysis of the proper motions and accelerations of the systemic features
(Herrnstein et al. 1999). Fifteen features were tracked over a period of
two years to an accuracy of 0.5--10~$\mu$as in relative position and
0.4 \kms\ in velocity. The distance  estimate is
based on simple geometric
considerations. The Keplerian curve of the high-velocity masers gives
the mass function $GM \sin^2 i/D$, where $i$ is the inclination of the
disk to the line of sight. The radius, $R$, of the systemic masers (in
angular units) is
determined from Equation (6), based on the slope of the velocity versus impact
parameter curve shown in Figure~7. This fixes the angular velocity,
$v_{\phi}$, of the systemic masers under the assumption that the orbits
are circular.
The expected accelerations and proper
motions of the systemic features are $v_\phi^2/R$ and $v_\phi/D$,  
respectively.
The assumption that the orbits are circular is reasonable on theoretical grounds
because of viscous relaxation and on observational grounds because the
continuum emission arises close to the center of symmetry of the
maser distribution. 

The distance to the 15 Cepheid variables in NGC\,4258 has been estimated to be
$8.1 \pm 0.8$~Mpc (Maoz {\it et al.} 1999). The statistical component of the
error is 0.4~Mpc and the systematic error associated with the calibration of 
the Cepheid distance scale is 0.7~Mpc. The discrepancy between the two distance
measurements to NGC\,4258 may have cosmological implications (Paczynski 1999).

\section{INTERESTING UNANSWERED QUESTIONS}

\noindent
1. What is the rate of radial inflow of material through the disk?
\nopagebreak

The accretion rate of material onto the black hole is an important
parameter that affects our understanding of radiation processes around the black hole.
The maser data
provide some information about the accretion rate.
Key issues are the long timescale needed for material to flow from the disk
to the black hole and the assumption that the masers trace all the disk material. 
It is useful to first estimate the mass of the disk. If we account for
systematic effects in addition to the random scatter of 3~\kms, the deviation from
Keplerian motion due to the finite
mass of the disk, $\Delta v_\phi$, is less than about 10~\kms\ over the radius of the disk. 
This limits the mass of the disk to less than about $2M\Delta v_\phi/v_\phi$, or about
$10^{6}$~\msun. The density of the molecular gas must be less than $\rho_c$ ($10^{10}$
hydrogen molecules per cubic centimeter). Since $H/R < 0.0025$, the upper limit on mass
is $10^5$~\msun. If, in addition, the disk is stable against the effects of
self-gravity (Toomre 1964; Binney \& Tremaine 1987), then the mass of the disk must 
be less than $M(H/R)$, or about $10^5$~\msun.

The mass accretion rate of a disk in steady state is given by
\begin{equation}
\dot{M} = 2\pi R \Sigma v_R~,
\end{equation}
%
where $\Sigma$ is the surface density of the disk, and $v_R$ is radial drift
velocity, which depends on the viscosity of the disk. Unfortunately, $v_R$ is
only weakly constrained by the observations (i.e., the possible difference between
the optical and radio systemic velocities) to be $<10$~\kms. This provides a
crude limit on the accretion rate of 100 \msunyr. To further constrain
the mass accretion rate requires an estimate of the viscosity of the disk.
In the standard model of 
a thin, viscous accretion disk, as formulated by Shakura and Sunyaev (1973), 
$v_R$ can be written (see Frank, King, \& Raine 1992) as
\begin{equation}
v_R = \alpha v_{\phi}{\left(H\over R\right)}^2~,
\end{equation}

\noindent
where $\alpha$  is the dimensionless viscosity parameter ($0 \le \alpha \le 1$).
The observational limit on the ratio $H/R$ implies that $v_R<0.006\alpha$~\kms.
With the limit on
mass given by the deviation from Keplerian motion, the accretion rate is less that 
$10^{-1}\alpha$~\msunyr. 
The infall time from the masing region is $R/v_R$, which
from Equation (10) can be written as
\begin{equation}
T \sim {1\over \alpha}{\left ({c\over c_S}\right )^2}{\left ({R_S\over c}\right)}%
{\left ({R\over R_S}\right )^{1\over 2}}~.
\end{equation}

\noindent
For NGC\_4258, with $\alpha = 0.1$ and c$_S = 2.5$ \kms , $T = 10^{16}$~s, or about
$3 \times 10^8$~yrs.

From the magnetic field limit and the assumption of
equipartition of  magnetic and thermal energy, the upper limit on $\dot{M}$
is also $10^{-1}\alpha$~\msunyr. If the maser density is the maximum allowable value,
$\rho_c$, and the
maser traces all the material in the disk, then the limit on disk thickness
leads to an upper limit on $\dot{M}$ of $10^{-2}\alpha$~\msunyr. Detailed theoretical modeling
can give estimates for the accretion rate. For example, a model in which the
cause of the outer radial cutoff in maser emission is attributed
to the transition from molecular to atomic gas leads to an
estimate of $10^{-4}\alpha$~\msunyr\ (Neufeld \& Maloney 1995). 
Gammie, Narayan, \& Blandford (1999) favor an accretion rate of $10^{-1}\alpha$~\msunyr,
based on an analysis of the continuum radiation spectrum.

If the accretion rate is high, then the relative weakness of the continuum 
radiation may be due to the process of advection (Gammie, Narayan, \& Blandford 1999). On the
other hand, if the accretion rate is low, then the weak emission is due to the
dearth of infalling material. In this case the gravitational power in the 
accretion flow may be insufficient to power the jets.

\noindent
2. What is the form and origin of the warp? 
\nopagebreak

The form of the warp is difficult to determine precisely, because the
filling factor of the masers in the disk is so small. Better measurements
of the positions and directions of motion of the high-velocity features are
key to defining the warp more accurately.

The cause of the warp is unknown, but several suggestions have been put
forward. Papaloizou, Terquem, \& Lin (1998) show that the warp could be
produced by a binary companion orbiting outside the maser disk. Its mass
would need to be comparable to the mass of the disk ($< 10^6$~\Mdot ).
Alternatively, radiation pressure from the central source will produce
torques on a slightly warped disk and will cause the warp to grow
(Maloney, Begelman, \& Pringle 1996). Finally, it is conceivable that
in the absence of other torques, the observed warp is due to the
Lense-Thirring effect. A maximally rotating black hole
will cause a precession of a nonaligned orbit (weak field limit) of
\begin{equation}
{\Omega}_{LT} = {{2G^2M^2}\over {c^3R^3}}~,
\end{equation}

\noindent
which can be rewritten in terms of the Schwarzschild radius as

\begin{equation}
{\Omega}_{LT} = {1\over 2}{c\over R_S}{\left ({R_S\over R}\right )^3}~.
\end{equation}

\noindent
At the inner radius of the disk ($R/R_S = 40,000$), the precession amounts
to $3 \times 10^{-17}$~s$^{-1}$. This precession is very small but 
might be significant over the lifetime of the disk. Equation~(11) suggests that the
lifetime might be $10^{16}$~s, which would produce a
differential precession of about 10\degrees~ across the radius of the disk. 
If the axis of the disk is inclined to the axis
of the black hole, then the viscosity of
the disk is expected to twist the plane of the innermost part of the disk to the equatorial
plane of the black hole (Bardeen \& Petterson 1975; Kumar \& Pringle, 1985).

\noindent
3. Do the water masers trace the whole disk?
\nopagebreak

The inner and outer radii of the observed masers are undoubtedly due to
excitation conditions in the maser. In the vertical direction it is also possible that the
masers form in a thin region within a thicker disk with atomic and ionized components. 
It has also been proposed that the high-velocity features are not indicative of 
a warped disk but trace material that has been blown off a flat disk (Kartje, Konigl, \& 
Elitzur 1999). These proposals are difficult to test.

\begin{figure}[t]
\plotfiddle{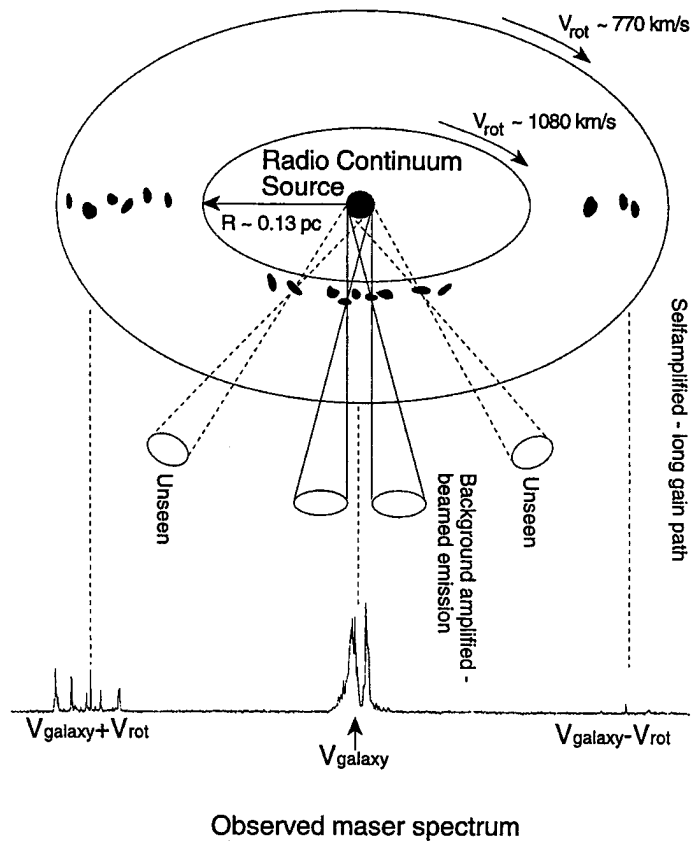}{3.6in}{0}{100}{100}{-310}{-224}
\caption{\setlength{\baselineskip}{8 pt}
A cartoon of the molecular accretion disk in NGC\_4258. The period of 
rotation at the inner edge of the disk is about 800 years. A particular
systemic maser is visible  for about 10 years as its
radiation beam, estimated to have a width of about 8\degrees , sweeps
over the earth. The high-velocity features
may be visible for a substantially longer time. (Adapted from 
Greenhill et al. 1995a)}
\end{figure}

\noindent
4. What are the physical properties  of the maser spots?
\nopagebreak

The spectrum of the maser has many discrete peaks which correspond to
spots of maser emission on the sky. The success of measuring proper motions and 
accelerations of these masers suggests that they correspond to discrete
condensations or density-enhanced regions in the disk. A cartoon of the
blobs in a disk is shown in Figure~11. The blobs in front of the
black hole may be visible because they amplify emission from the
central region. No masers have been seen on the backside of the disk. 
On the other hand, the high-velocity masers have no continuum emission to
amplify, and we may only see the ones near the midline, where the 
gradient in the line-of-sight velocity is small. Blobs in the rest of
the disk may be radiating in directions away from the earth. The
clumpiness of the medium allows us to track 
the individual masers. If the appearance of spots is due to blobs, or
density enhancements in the disk, then
the minimum gradient condition would not seem to be necessary. However, intense
high-velocity maser spots may occur when blobs at the same velocity
line up to form two-stage masers (Deguchi \& Watson 1989).
This situation forms a highly
beamed maser, like the filamentary maser described in Section~2. The probability of realizing
this situation is greatest along the midline, where the velocity gradient is
smallest.  All evidence suggests that the masers arise from discrete
physical condensations. There have been several suggestions that
the apparent motions of the maser spots may be due to a phase effect
(e.g., a spiral density wave
moving through the disk, Maoz \& McKee 1998), but there is no
observational evidence for this.

\noindent


\begin{table}[b]
\begin{center}
\begin{tabular}{lcccc}
\multicolumn{5}{c}{TABLE 2}\\[1ex]
\multicolumn{5}{c}{\sc Known AGN with Water Masers}\\[1ex]
\hline\hline\noalign{\vspace{1ex}}
            & Distance   & Flux Density  & Milliarcsecond        &         \\
Galaxy & (Mpc)      & (Jy)          & Structure  & Disk    \\[.5ex]
\hline\noalign{\vspace{1ex}}
M\_51 	     & 3   & 0.2  & \nodata & \nodata  	\\
NGC\_4945    & 3.7 & 4    & yes & yes 	\\
Circinus     & 4   & 4    & yes & yes 	\\
NGC\_4258    & 7.2 & 4    & yes & yes 	\\
NGC\_1386    & 12  & 0.9  & yes & maybe \\
NGC\_3079    & 16  & 6    & yes & maybe \\
NGC\_1068    & 16  & 0.6  & yes & yes 	\\
NGC\_1052    & 20  & 0.3  & yes & no 	\\
NGC\_613     & 20  & 0.1? & \nodata & \nodata	\\
NGC\_5506    & 24  & 0.6  & \nodata & \nodata 	\\
NGC\_5347    & 32  & 0.1  & \nodata & \nodata	\\
NGC\_3735    & 36  & 0.2  & \nodata & \nodata 	\\
IC\_2560     & 38  & 0.4  & yes	    & \nodata 	\\
NGC\_2639    & 44  & 0.1  & \nodata & \nodata 	\\
NGC\_5793    & 50  & 0.4  & \nodata & \nodata 	\\
ESO\_103-G035 & 53  & 0.7 & \nodata & \nodata	\\
Mrk\_1210    & 54  & 0.2  & \nodata & \nodata 	\\
IRAS F01063-8034 & 57 & 0.2 & \nodata & \nodata \\
Mrk\_1       & 65  & 0.1  & \nodata & \nodata	\\
NGC\_315     & 66  & 0.05 & \nodata & \nodata 	\\
IC\_1481     & 83  & 0.4  & \nodata & \nodata 	\\
IRAS F22265-1826  & 100 & 0.3  & yes & no 	\\
\noalign{\vspace{.5ex}}\hline \noalign{\vspace{1.5 ex}}
\multicolumn{5}{l}{\parbox{4.45in}{\parindent=0pt \baselineskip=10pt \raggedright
		          {\sc Note}--- Extragalactic masers outside AGN are found in
		          NGC\_253, M\_82, M\_33, IC\_342, LMC, and SMC.}}
\end{tabular}
\end{center}
\end{table}

\section{MASERS IN OTHER AGN}

At this time (early 1999), 22 masers have been detected among about 700 galaxies
searched (e.g., Braatz, Wilson, \& Henkel 1997).  A list of these galaxies with
masers is given in Table 2. The yield rate of detections is only about 3 percent.
The major reason for this paucity is probably that the maser disks can
only be seen if they are edge-on to the line of sight. If the
typical beam angle, $\beta$, is 8\degrees, as in NGC\_4258, then the probability
of seeing a maser is about equal
to $\sin \beta,$ or 8 percent. Braatz, Wilson, \& Henkel (1997) have
shown that most of the known masers are associated with Seyfert II galaxies
or LINERs where the accretion disks are thought to be edge-on to the earth.


\begin{table}[b]
\begin{center}
\begin{tabular}{l@{}ccrcccl}
\multicolumn{8}{c}{TABLE 3}\\[1ex]
\multicolumn{8}{c}{\sc Water Masers with Resolved Structure}\\[1ex]
\hline\hline\noalign{\vspace{2ex}}
\multicolumn{8}{c}{Masers Without Obvious Disk Structure}\\[1ex]
\hline\noalign{\vspace{1ex}}
       && $D$     & \omit{\hfil $v_0$\hfil}  & $\Delta v$ & $\Delta R$ &         &           \\
Galaxy && \sm Mpc & \omit{\hfil \sm \kms\hfil}  & \sm \kms   & \sm pc     & Comment & Reference \\
\hline\noalign{\vspace{1ex}}
IRAS\_22265 (S0) && 100 & \omit{\hfil 7570\hfil} & 150   & 2.4 & messy      & Greenhill et al. 1999a\\
NGC\_1052 (E4)   && \z 20  & \omit{\hfil 1490\hfil} & 100   & 0.6 & ``jet''    & Claussen et al. 1998\\
IC\_2560         && \z 38  & \omit{\hfil 2900\hfil} & \z 30 & 0.2 & velocity   & Nakai et al. 1998  \\
                &&     &                        &       &     & gradient   &                \\
\noalign{\vspace{.5ex}}\hline
\hline
\noalign{\vspace{2ex}}
\multicolumn{8}{c}{Masers With Disk Structure}\\[1ex]
\hline\noalign{\vspace{1ex}}
       & $D$ & $v_{\phi}$ & \omit{\hfil $R_i$/$R_o$\hfil} & $M$ & $\rho$ & $L_x$ &        \\
Galaxy & \sm Mpc & \sm \kms & \omit{\hfil \sm pc\hfil} & \sm 10$^6$ M$_{\odot}$
	& \sm  10$^7$ M$_{\odot}\,$pc$^{-3}$ & \sm 10$^{42}$ erg$\,$s$^{-1}$ & Reference \\ 
\hline\noalign{\vspace{1ex}}
NGC\_4258 & \z 7 & 1100    & 0.13/0.26   &   35 & 400 & 0.04 & Miyoshi et al. 1995\\
NGC\_1068 & 15   & \z  330 & 0.6/1.2\z   &   17 &   3 & 40   & Greenhill \& Gwinn 1997\\
Circinus & \z 4 & \z  230 & 0.08/0.8\z  & \z 1 &  40 & 40   & Greenhill et al. 1999b\\
NGC\_4945 & \z 4 & \z  150 & 0.2/0.4\z   & \z 1 &   2 &  1   & Greenhill et al. 1997\\
NGC\_1386 & 12   & \z  100 & --/0.7\z    & \z 2 &   4 & 0.02 & Braatz et al. 1999\\
NGC\_3079 & 16   & \z  150 & --/1.0\z    & \z 1 & 0.2 & 0.02 & Trotter et al. 1998,\\
         &      &         &             &      &     &      & Satoh et al. 1998 \\
\noalign{\vspace{.5ex}}\hline
\noalign{\vspace{1.5ex}}
\multicolumn{8}{l}{\parbox{6.5 in}{\parindent=0pt \baselineskip=10pt \raggedright
$D$ = distance,
$v_0$ = systemic velocity, $\Delta v$ = velocity range, $\Delta R$ = linear extent,
$v_{\phi}$ = rotational velocity, $R_i$/$R_o$ = inner/outer radius of disk, 
$M$ = central mass, $\rho$ = central mass density, $L_x$ = X-ray luminosity.}}\\
\noalign{\vspace{1ex}}
\end{tabular}
\end{center}
\end{table}

It is difficult to make VLBI measurements on masers weaker than about 0.5 Jy because
of the need to detect the maser within the coherence time of the interferometer.
Nine masers have been studied with VLBI. Four of these
show strong evidence of disk structure, and two more show 
probable disk structure. The properties of these masers
are listed in Table 3. Unfortunately, none of these masers 
show the simple, well-defined structure that would make them useful for
precise study of the physical properties of accretion disks around black holes. 

\section{SUMMARY}

The measurements of the positions and velocities  of the masers in the nucleus
of NGC\_4258 offer compelling evidence for the existence of a supermassive black
hole and provide the first direct image of an accretion disk within $10^5$~R$_S$
of the black hole. Much more can be learned from this system. A measurement of
the disk thickness is important and may require higher signal-to-noise ratios
than are achievable currently or VLBI measurements from space. Measurement of
the continuum spectrum from the central region is very important to the
understanding of the radiation process. Detection of radio emission would require
instruments of higher sensitivity. Continued
measurements over time of the positions and velocities of the masers will refine
the estimates of their proper motions and accelerations, and this will better
define the shape of the disk. It is even conceivable that the radial drift
velocity will be detected. This work will benefit immensely from new instruments
that are in the planning stage for centimeter wavelength radio astronomy. These
include the enhanced Very Large Array, the Square Kilometer Array, and space
VLBI missions such as ARISE. 

We thank Adam Trotter and Ann Bragg for helpful discussions.

\begin{center}
{\bf REFERENCES}
\end{center}

\setlength{\parindent}{0pt}

Antonucci, R.~R.~J. \& Miller, J.~S. 1985, {\it Astrophys. J.}, {\bf 297}, 621--632.

Bardeen, J.~M. \& Petterson, J.~A. 1975, {\it Astrophys. J.}, {\bf 195}, L65--L67.

Begelman, M.~C. \& Rees, M.~J. 1978, {\it Mon. Not. R. Astr. Soc.}, {\bf 185}, 847--859.

Binney, J. \& Tremaine, S. 1987, {\it Galactic Dynamics} (Princeton: Princeton Univ. Press).

Blandford, R.~D. \& Gehrels, N. 1999, {\it Physics Today}, {\bf 52}, 40--46.

Blandford, R.~D. \& Rees, M. 1992, in {\it Testing the AGN Paradigm}, ed. S.~S. Holt,
S.~G. Neff, \& C.~M. Urry (New York: American Inst. of Physics), 3--19.

Blandford, R.~D. \& Znajek, R.~L. 1977, {\it Mon. Not. R. Astr. Soc.}, {\bf 179}, 433--456.

Bragg, A.~E., Greenhill, L.~J., Moran, J.~M., \& Henkel, C. 1999, {\it Astrophys. J.}, submitted.

Braatz, J.~A., Wilson, A.~S., \& Henkel, C. 1997, {\it Astrophys. J. Supp.}, {\bf 110}, 321--346.

Braatz, J.~A., et. al. 1999, {\it Astrophys. J.}, in preparation.

Bromley, B.~C., Miller, W.~A., \& Pariev, V.~I. 1998, {\it Nature}, {\bf 391}, 54--56.

Cecil, G., Wilson, A.~S., \& Tully, R.~B. 1992, {\it Astrophys. J.}, {\bf 390}, 365--377.

Churchwell, E., Witzel, A., Huchtmeier, W., Pauliny-Toth, I., Roland, J., \& Sieber, W.
1977, {\it Astr. Astrophys.}, {\bf 54}, 969--971.

Claussen, M.~J., Diamond, P.~J., Braatz, J.~A., Wilson, A.~S., \&
Henkel, C. 1998, {\it Astrophys. J.}, {\bf 500}, L129--L132.

Claussen, M.~J., Heiligman, G.~M., \& Lo, K.~Y. 1984, {\it Nature}, {\bf 310}, 298--300.

Claussen, M.~J. \& Lo, K.-Y. 1986, {\it Astrophys. J.}, {\bf 308}, 592--599.

Deguchi, S. \& Watson, W.~D. 1989, {\it Astrophys. J.}, {\bf 340}, L17--L20.

dos Santos, P.~M. \& Lepine, J.~R.~D. 1979, {\it Nature}, {\bf 278}, 34--35.

Elitzur, M. 1992, {\it Astronomical Masers} (Dordrecht: Kluwer).

Faber, S.~M. 1999, in Proceedings of the 32nd COSPAR Meeting, {\it The AGN-Galaxy
Connection,} ed. H~R. Schmitt, L.~C. Ho, \& A.~L. Kinney (Advances in Space Research), in press.

Frank, J., King, A., \& Raine, D. 1992, {\it Accretion Power in Astrophysics}
(Cambridge: Cambridge Univ. Press).

Gammie, C.~F., Narayan, R., \& Blandford, R. 1999, {\it Astrophys. J.},
{\bf 516}, 177--186.

Gardner, F.~F. \& Whiteoak, J.~B. 1982, {\it Mon. Not. R. Astr. Soc.},
{\bf 201}, 13p--15p.

Genzel, R., Eckart, A., Ott, T., \& Eisenhauer, F. 1997, {\it Mon. Not. R.
Astr. Soc.,} {\bf 291}, 219--234.

Ghez, A.~M., Klein, B.~L., Morris, M., \& Becklin, E.~E. 1998, {\it Astrophys. J.},
{\bf 509}, 678--686.

Greenhill, L.~J., et al. 1999a, in preparation.

Greenhill, L.~J., et al. 1999b, in preparation.

Greenhill, L.~J. \& Gwinn, C.~R. 1997, {\it Astrophys. Space Sci.}, {\bf 248},
261--267.

Greenhill, L.~J., Henkel, C., Becker, R., Wilson, T.~L., \& Wouterloot, J.~G.~A.
1995b, {\it Astr. Astrophys.}, {\bf 304}, 21--33.

Greenhill, L.~J., Jiang, D.~R., Moran, J.~M., Reid, M.~J., Lo, K.~Y., \& Claussen, M.~J.
1995a, {\it Astrophys. J.}, {\bf 440}, 619--627.

Greenhill, L.~J., Moran, J.~M., \& Herrnstein, J.~R. 1997, {\it Astrophys. J.},
{\bf 481}, L23--L26. 

Haschick, A.~D., Baan, W.~A., \& Peng, E.~W. 1994, {\it Astrophys. J.}, {\bf 437},
L35--L38.

Heckman, T.~M. 1980, {\it Astr. Astrophys.}, {\bf 87}, 152--164.

Herrnstein, J.~R. 1996, Ph.D. thesis, Harvard University.

Herrnstein, J.~R., Greenhill, L.~J., \& Moran, J.~M. 1996, {\it Astrophys. J.}, {\bf 468},
L17--L20.

Herrnstein, J.~R., Greenhill, L.~J., Moran, J.~M., Diamond, P.~J., Inoue, M.,
Nakai, N., \& Miyoshi, M. 1998b, {\it Astrophys. J.}, {\bf 497}, L69--L73.

Herrnstein, J.~R., Moran, J.~M., Greenhill, L.~J., Blackman, E.~G., \& Diamond, P.~J.
1998a, {\it Astrophys. J.}, {\bf 508}, 243--247.

Herrnstein, J.~R., Moran, J.~M., Greenhill, L.~J., Diamond, P.~J., Inoue, M., Nakai, N., 
Miyoshi, M., Henkel, C., \& Riess, A.  1999, {\it Nature}, {\bf 400}, 539--541.

Herrnstein, J.~R., Moran, J.~M., Greenhill, L.~J., Diamond, P.~J.,
Miyoshi, M., Nakai, N., \& Inoue, M. 1997, {\it Astrophys. J.}, {\bf 475}, L17--L20.

Ho, L.~C. 1999, in {\it Observational Evidence for Black Holes in the Universe},
ed. S.~K. Chakrabarti (Dordrecht: Kluwer), 157--186.

Kartje, J.~F., Konigl, A., \& Elitzur, M. 1999, {\it Astrophys. J.}, {\bf 513}, 180--196.

Kumar, S. \& Pringle, J.~E. 1985, {\it Mon. Not. R. Astr. Soc.}, {\bf 213}, 435--442.

Kormendy, J. \& Richstone, D. 1995, {\it Ann. Rev. Astr. Astrophys.}, {\bf 33}, 581--624.

Maloney, P.~R., Begelman, M.~C., \& Pringle, J.~E. 1996, {\it Astrophys. J.},
{\bf 472}, 582--587.

Maoz, E. 1995, {\it Astrophys. J.}, {\bf 455}, L131--L134.

Maoz, E., {\it et al.} 1999, {\it Nature,} {\bf 401}, 351--354. 

Maoz, E. \& McKee, C. 1998, {\it Astrophys. J.}, {\bf 494}, 218--235.

Miyoshi, M. 1999, in {\it Observational Evidence for Black Holes in the Universe},
ed. S.~K. Chakrabarti (Dordrecht: Kluwer), 141--156.

Miyoshi, M., Moran, J.~M., Herrnstein, J.~R., Greenhill, L.~J., Nakai, N., Diamond, P.~D.,
\& Inoue, M. 1995, {\it Nature}, {\bf 373}, 127--129.

Moran, J. 1981, {\it Bull. Am. Astron. Soc.}, {\bf 13}, 508.

Moran, J.~M., Greenhill, L.~J., Herrnstein, J.~R., Diamond, P.~D., Miyoshi, M.,
Nakai, N., \& Inoue, M. 1995, {\it Proc. Nat. Acad. Sci.}, USA, {\bf 92}, 11427--11433.

Nakai, N., Inoue, M., Hagiwara, Y., Miyoshi, M., \& Diamond, P.~J. 1998,
in {\it Radio Emission from Galactic and Extragalactic Compact Sources},
ed. J.~A. Zensus, G.~B. Taylor, \& J.~M. Wrobel (San Francisco: ASP), 237--238.

Nakai, N., Inoue, M., Miyazawa, K., Miyoshi, M., \& Hall, P. 1995, {\it Pub. Astron.
Soc. Jap.}, {\bf 47}, 771--799.

Nakai, N., Inoue, M., \& Miyoshi, M. 1993, {\it Nature}, {\bf 361}, 45--47.

Neufeld, D. \& Maloney, P.~R. 1995, {\it Astrophys. J.}, {\bf 447}, L17--L20.

Paczynski, B. 1999, {\it Nature,} {\bf 401}, 331--332.

Papaloizou, J.~C.~B., Terquem, C., \& Lin, D.~N.~C. 1998, {\it Astrophys. J.}, {\bf 497}, 212--226.

Reid, M.~J. \& Moran, J.~M. 1988, in {\it Galactic and Extragalactic Radio Astronomy},
2d ed., ed. G.~L. Verschuur \& K.~I. Kellermann (New York: Springer-Verlag), 255--294.

Rees, M. 1998, in {\it Black Holes and Relativistic Stars}, ed. R.~M. Wald (Chicago: Univ.
Chicago Press), 79--101.

Salpeter, E.~E. 1964, {\it Astrophys. J.}, {\bf 140}, 796--800.

Satoh, S., Inoue, M., Nakai, N., Shibata, K.~M., Kameno, S., Migenes, V., \&
Diamond, P.~J. 1998, in {\it Highlights of Astronomy}, {\bf 11b}, ed. J. Andersen
(Dordrecht: Kluwer), 972--973.

Seyfert, C. 1943, {\it Astrophys. J.}, {\bf 97}, 28--40.

Shakura, N.~I. \& Sunyaev, R.~A. 1973, {\it Astr. Astrophys.}, {\bf 24}, 337--355.

Tanaka, Y., et al. 1995, {\it Nature}, {\bf 375}, 659--661. 

Thompson, A. ~R., Moran, J.~M., \& Swenson, G.~W. 1986, {\it Interferometry and 
Synthesis in Radio Astronomy} (New York: Wiley Interscience).

Toomre, A. 1964, {\it Astrophys. J.}, {\bf 139}, 1217--1238.

Trotter, A.~S., Greenhill, L.~J., Moran, J.~M., Reid, M.~J., Irwin, J.~A., \& Lo, K.~Y. 
1998, {\it Astrophys. J.}, {\bf 495}, 740--748.

Weaver, H., Williams, D.~R.~W., Dieter, N.~H., \& Lum, W.~T. 1965,
{\it Nature}, {\bf 208}, 29--31.

\eject

\begin{center}
{\bf QUESTIONS AND ANSWERS}
\end{center}

\begin{list}{}{\settowidth{\labelwidth}{Q:~~}
               \setlength{\labelsep}{0pt}
               \setlength{\leftmargin}{\labelwidth}
               \addtolength{\leftmargin}{\labelsep}
               \setlength{\itemsep}{12pt}
               \renewcommand{\makelabel}[1]{#1\hfill}}
\item [Q:~~] What other measurements can one do with masers to try to get
sensitivity to the general relativistic effects or to get information from
distances closer to the mass concentration?
\item [A:~~] We have searched for masers in NGC$\,$4258 that are closer to the central
mass than 40,000 Schwarzschild radii and have found none. The reason for
the absence of masers closer to the center is unclear. Masers are very sensitive
to local conditions such as temperature, density, and pump power,
which may depend on the geometry of the warp. Water masers arise in neutral
media at temperatures below a few thousand degrees, so it is unrealistic
to expect to find masers very close to the Schwarzschild radius.

One possible general relativity effect might be the deflection of radiation
from masers on the far side of the accretion disk (i.e., a distortion of the
image of the back side of the accretion disk). However, no such masers have
been found.
\end{list}

\end{document}